\documentclass[journal]{IEEEtran}
\ifCLASSINFOpdf
\else
\fi
\usepackage{cite}

\usepackage{graphicx}
\usepackage{subfigure}
\usepackage[outdir=./]{epstopdf}

\usepackage{bm}
\usepackage{amsfonts}
\usepackage{dsfont}
\usepackage{amssymb}

\usepackage{color}

\usepackage{amsmath}
\usepackage{flushend}
\usepackage{algorithm}
\usepackage{algorithmic}

\usepackage{booktabs}

\ifCLASSOPTIONcompsoc
 \usepackage[caption=false,font=normalsize,labelfont=sf,textfont=sf]{subfig}
\else
 \usepackage[caption=false,font=footnotesize]{subfig}
\fi
\usepackage{mathrsfs}

\usepackage{stfloats}
\fnbelowfloat
\begin{document}
%
\title{A Convex Hull Based Approach for MIMO Radar Waveform Design with Quantized Phases}
%
%
%

\author{Chenglin~Ren, 
        Zhaohui Ma,
        Fan~Liu,~\IEEEmembership{Student Member,~IEEE,}
        Weichao~Pi, 
        and~Jianming~Zhou 
\thanks{C. Ren, Z. Ma, F. Liu, W. Pi and J. Zhou are with the School of Information and Electronics, Beijing Institute of Technology, Beijing, 100081, China (e-mail: rencl3@bit.edu.cn; mzh\_19890822@bit.edu.cn; liufan92@bit.edu.cn; peao\_kelvin@163.com; zhoujm@bit.edu.cn)}
}

\maketitle

\begin{abstract}In this letter, we focus on designing constant-modulus waveform with discrete phases for the multi-input multi-output (MIMO) radar, where the signal-to-interference-plus-noise ratio (SINR) is maximized in the presence of both the signal-dependent clutter and the noise. Given the NP-hardness of the formulated problem, we propose to relax the original optimization as a sequence of continuous quadratic programming (QP) subproblems by use of the convex hull of the discrete feasible region, which yields approximated solutions with much lower computational costs. Finally, we assess the effectiveness of the proposed waveform design approach by numerical simulations.

\end{abstract}

\begin{IEEEkeywords}
waveform design, discrete phases, MIMO radar, quadratic programming, convex hull
\end{IEEEkeywords}

%
\IEEEpeerreviewmaketitle

\section{Introduction}
%
%
%
%
\IEEEPARstart {A}{ccording} to the configuration of the antennas, the MIMO radar is studied in two types: distributed MIMO radar \cite{1597550,4408448} and colocated MIMO radar \cite{1399143,4350230}. Compared with traditional phased array radar, MIMO radar offers extra degrees of freedom by allowing individual waveforms to be transmitted at each antenna, and therefore achieves enhanced capabilities \cite{4156404,4400830,4655353}. In addition, a well designed waveform can significantly improve the SINR \cite{6136800,7140824} and probability of detection \cite{259642}.

The problem of waveform design for MIMO radar has gained considerable attention in recent years \cite{5263002,6194367,6563125,6649991,7126162,7450660,arxiv1805}. In practical scenarios, waveforms with constant modulus or low peak-to-average-power ratio (PAPR) properties are needed to avoid signal distortions, as the non-linear power amplifiers are typically operated at the saturation region. Meanwhile, the similarity constraint (SC) is also enforced, which employs a reference waveform as the benchmark and allows the optimized waveform to share some of the ambiguity properties of the reference waveform. Relying on the semidefinite relaxation (SDR) and rank-one decomposition techniques, the authors of \cite{6649991} introduce the sequential optimization procedures to maximize the SINR accounting for the constant modulus constraint (CMC) and the SC for the case of continuous signal phase. In \cite{7450660}, a novel successive QCQP refinement (SQR) algorithm is developed for the same optimization problem in \cite{6649991}, which involves solving a sequence of convex quadratically constrained quadratic programming (QCQP) subproblems. Furthermore, by customizing the projection procedure for the convex QCQP subproblems, the Accelerated Gradient Projection (AGP) is proposed in \cite{arxiv1805}, which shares a comparable performance with the SQR method while notably reducing the computational complexity. However, the works aforementioned only address the continuous phase case. Given the extensive use of digital phase shifters in many radar systems, we consider in this letter the MIMO radar waveform design for the case of quantized phases. Pioneered by \cite{4668417}, the cases of both continuous and discrete phases are considered. Based on the method of semidefinite program (SDP) relaxation and randomization, the approximated solutions to the quadratic optimization problem subject to the CMC and the SC are presented. In \cite{5732713}, the optimization problem under the PAPR constraint and energy constraint is resolved through the same techniques.


In this letter, we optimize the SINR of the radar under both CMC and SC, where the phases of the designed signals are drawn from a discrete alphabet. By introducing the convex hull of the finite feasible points, a novel continuous approximation method (CAM) is proposed to relax the original problem as a sequence of convex QP subproblems, which can be solved via standard numerical tools. By doing so, the solution of the original discrete problem can be then obtained by a simple quantization procedure. Numerical results show that the performance of the  proposed approach can approximate that of the continuous waveform design.

\section{System Model}
We consider a colocated narrow band MIMO radar system equipped with ${N_T}$ transmit antennas and ${N_R}$ receive antennas, where each antenna emits or receives $N$ samples. The receive waveform is given as \cite{6649991}
\begin{equation}\label{1}
{\mathbf{r}} = {\alpha _0}{\mathbf{A}}({\theta _0}){\mathbf{s}} + \sum\limits_{m = 1}^M {{\alpha _m}{\mathbf{A}}({\theta _m}){\mathbf{s}} + {\mathbf{n}}}
\end{equation}
where ${\mathbf{r}} = {[{\mathbf{r}}_1^T, \ldots ,{\mathbf{r}}_N^T]^T} \in \mathbb{C}^{N_RN \times 1}$ with ${{\mathbf{r}}_n} \in {\mathbb{C}^{{N_R} \times 1}}$, $n = 1, \ldots ,N$ being the $n$-th snapshot across the ${N_R}$ receive antennas, ${\mathbf{s}} = {[{\mathbf{s}}_1^T, \ldots ,{\mathbf{s}}_N^T]^T} \in \mathbb{C}^{N_TN \times 1}$ stands for the transmit waveform with ${{\mathbf{s}}_n} \in {\mathbb{C}^{{N_T} \times 1}}$ being the $n$-th snapshot across the ${N_T}$ transmit antennas, ${\mathbf{n}} \in {\mathbb{C}^{N_RN \times 1}} \sim \mathcal{CN}(0,\sigma _n^2{\mathbf{I}})$  denotes circular white Gaussian noise, ${\alpha _0}$ and ${\alpha _m}$ represent the complex amplitudes of the target and the $m$-th interference source, ${\theta _0}$ and ${\theta _m}$ represent the angles of the target and the $m$-th interference source, respectively, and ${\mathbf{A}}(\theta )$ stands for the steering matrix of a Uniform Linear Array (ULA) with half-wavelength separation between the antennas, which is given as
\begin{equation}\label{2}
{\mathbf{A}}(\theta ) = {{\mathbf{I}}_N} \otimes [{{\mathbf{a}}_r}(\theta ){{\mathbf{a}}_t}{(\theta )^T}]
\end{equation}
where ${{\mathbf{I}}_N}$ is the $N \times N$ identity matrix, $\otimes$ denotes the Kronecker product, ${{\mathbf{a}}_t}$ and ${{\mathbf{a}}_r}$ stand for the transmit and the receive steering vectors, respectively.

Aiming for maximizing the output SINR of the radar, we jointly design the transmit waveform and the receive filter. Without loss of generality, we assume that a linear Finite Impulse Response (FIR) filter ${\mathbf{f}} \in \mathbb{C}^{N_RN \times 1}$ is employed to process the receive echo wave. The output of the filter $r$ is given as
\begin{equation}\label{3}
r = {{\mathbf{f}}^H}{\mathbf{r}} = {\alpha _0}{{\mathbf{f}}^H}{\mathbf{A}}({\theta _0}){\mathbf{s}} + \sum\limits_{m = 1}^M {{\alpha _m}{{\mathbf{f}}^H}{\mathbf{A}}({\theta _m}){\mathbf{s}}}  + {{\mathbf{f}}^H}{\mathbf{n}}
\end{equation}
where ${( \cdot )^H}$ denotes the Hermitian transpose. As a consequence, the output SINR can be expressed as
\begin{equation}\label{4}
\text{SINR} = \frac{{\sigma {{\left| {{{\mathbf{f}}^H}{\mathbf{A}}({\theta _0}){\mathbf{s}}} \right|}^2}}}{{{{\mathbf{f}}^H}{\mathbf{\hat T}}({\mathbf{s}}){\mathbf{f}} + {{\mathbf{f}}^H}{\mathbf{f}}} }
\end{equation}
where $\sigma  = E[{\left| {{\alpha _0}} \right|^2}]/\sigma _n^2$ with $E[ \cdot ]$ denoting the statistical expectation, and
\begin{equation}\label{5}
{\mathbf{\hat T}}({\mathbf{s}})  = \sum\limits_{m = 1}^M {{I_m}{\mathbf{A}}({\theta _m}){\mathbf{s}}{{\mathbf{s}}^H}{{\mathbf{A}}^H}({\theta _m})}
\end{equation}
where ${I_m} = E[{\left| {{\alpha _m}} \right|^2}]/\sigma _n^2$.

\section{Problem Formulation}
Based on the system model analyzed above, the waveform design with continuous phase is firstly considered, where the CMC and the SC aforementioned have been imposed. Thus, the maximization of the SINR in ($4$) can be formulated as
\begin{equation}\label{6}
\begin{gathered}
  \mathop {\max }\limits_{{\mathbf{f}},{\mathbf{s}}}\;\;\; \frac{{\sigma {{\left| {{{\mathbf{f}}^H}{\mathbf{A}}({\theta _0}){\mathbf{s}}} \right|}^2}}}{{{{\mathbf{f}}^H}{\mathbf{\hat T}}({\mathbf{s}}){\mathbf{f}} + {{\mathbf{f}}^H}{\mathbf{f}}}} \hfill \\
  s.t.\;\;\;\;\; \left| {s(k)} \right| = 1/\sqrt {{N_T}N} \hfill \\
  \;\;\;\;\;\;\;\;\;\; {\left\| {{\mathbf{s}} - {{\mathbf{s}}_0}} \right\|_\infty } \le \varepsilon  \hfill \\
\end{gathered}
\end{equation}
where ${{s}}(k)$ stands for the $k$-th entry of ${\mathbf{s}}$, $k = 1, \ldots ,{N_T}N$, ${\left\|  \cdot  \right\|_\infty }$ denotes the infinity norm, ${{\mathbf{s}}_0}$ represents the reference waveform, and $\varepsilon$ ($0 \le \varepsilon  \le 2$) is a preset parameter to control the similarity degree between ${\mathbf{s}}$ and ${{\mathbf{s}}_0}$. In addition, by taking into account the CMC, the SC can be rewritten as
\begin{equation}\label{7}
\arg {{s}}(k) \in [{\omega _k},{\omega _k} + \delta ]
\end{equation}
where ${\omega _k}$ and ${\delta}$ are respectively given as
\begin{equation}\label{8}
\begin{gathered}
  {\omega _k} = \arg {{{s}}_0}(k) - \arccos(1 - {\varepsilon ^2}/2) \hfill \\
  \delta = 2\arccos (1 - {\varepsilon ^2}/2) \hfill \\
\end{gathered}
\end{equation}
where ${{s_0}}(k)$ is the $k$-th entry of ${\mathbf{s}_0}$. Noting that there is no constraints on ${\mathbf{f}}$ as analyzed in \cite{6649991}, the maximization problem of ($6$) can be equivalent to
\begin{equation}\label{9}
\begin{gathered}
  \mathop {\max }\limits_{\mathbf{s}} \;\;\;{{\mathbf{s}}^H}{\mathbf{Y}}({\mathbf{s}}){\mathbf{s}} \hfill \\
  s.t.\;\;\;\;\;\left| {s(k)} \right| = 1/\sqrt {{N_T}N}  \hfill \\
  \;\;\;\;\;\;\;\;\;\;\;\arg s(k) \in [{\omega _k},{\omega _k} + \delta ] \hfill \\
\end{gathered}
\end{equation}
where ${\mathbf{Y}}({\mathbf{s}})$ is a positive-semidefinite matrix, which is given as
\begin{figure}[t]
  \centering
  \includegraphics[width=0.7\columnwidth]{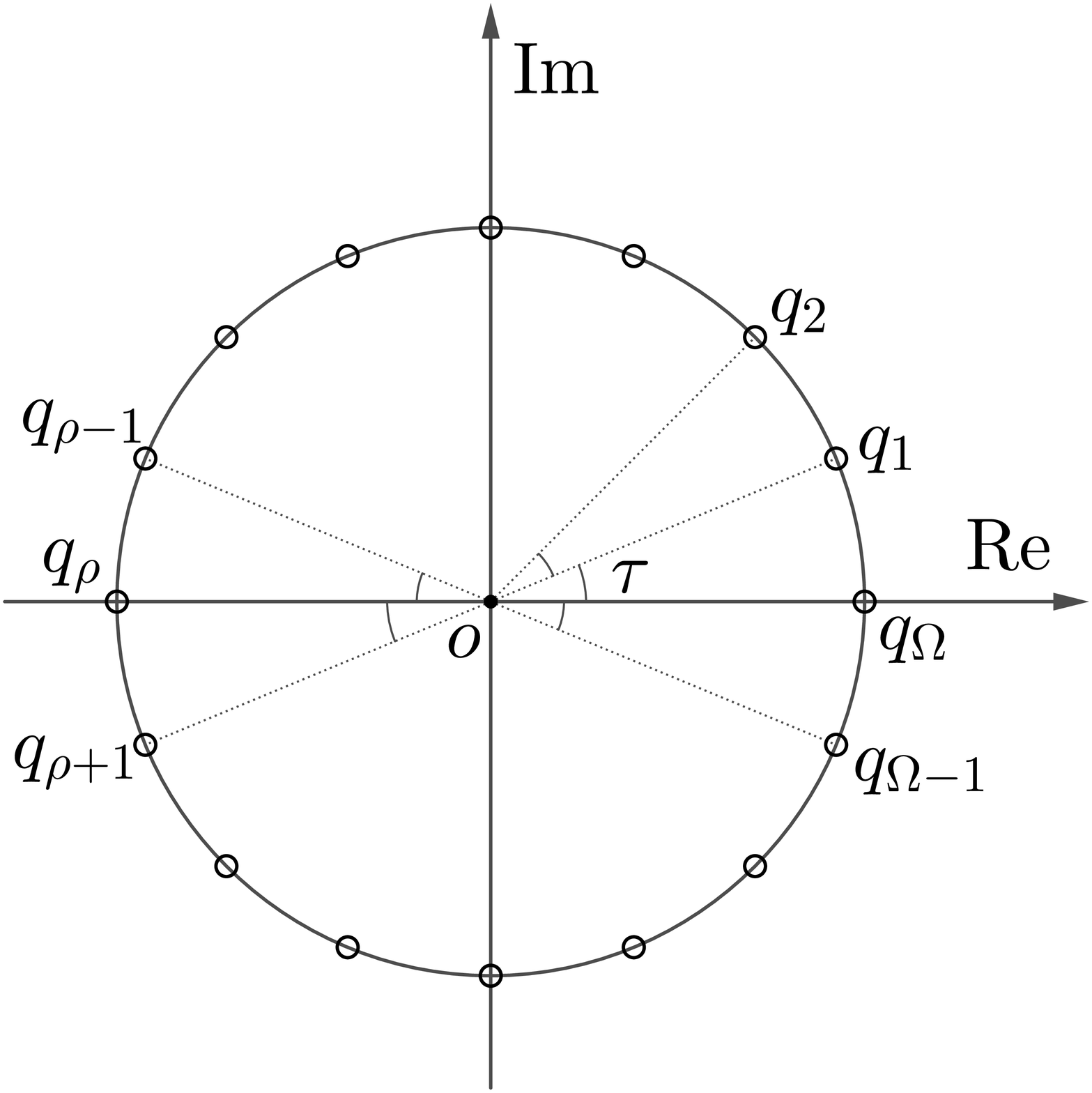}
  \caption{The constellation points with $\Omega = 16$ for $k$-th dimension.}
  \label{1}
\end{figure}
\begin{equation}\label{10}
{\mathbf{Y}}({\mathbf{s}}) = {{\mathbf{A}}^H}({\theta _0}){[{\mathbf{\hat T}}({\mathbf{s}}) + {\mathbf{I}}]^{ - 1}}{\mathbf{A}}({\theta _0}).
\end{equation}

According to \cite{4383615}, we can obtain a suboptimal SINR by assuming ${\mathbf{Y}} = {\mathbf{Y}}({\mathbf{s}})$ with a fixed $\mathbf{s}$ and optimizing $\mathbf{s}$ with the new $\mathbf{Y}$ iteratively, resulting in a sequence of subproblems as follows
\begin{equation}\label{11}
\begin{gathered}
  \mathop {\max }\limits_{\mathbf{s}} \;\;\;{{\mathbf{s}}^H}{\mathbf{Y}}{\mathbf{s}} \hfill \\
  s.t.\;\;\;\;\;\left| {s(k)} \right| = 1/\sqrt {{N_T}N}  \hfill \\
  \;\;\;\;\;\;\;\;\;\;\;\arg s(k) \in [{\omega _k},{\omega _k} + \delta ]. \hfill \\
\end{gathered}
\end{equation}
We consider the subproblem as ($11$) for each iteration in the sequel. In addition, let ${\mathbf{q}} \in \mathbb{C}^{N_TN \times 1}$ be the quantized transmit waveform, and $q(k)$ be the $k$-th entry of ${\mathbf{q}}$. As a consequence, the maximization of the SINR in ($11$) for the continuous phase case can be quantized as
\begin{equation}\label{12}
\begin{gathered}
  \mathop {\max }\limits_{\mathbf{q}}\;\;\; {{\mathbf{q}}^H}{\mathbf{Y}}{\mathbf{q}} \hfill \\
  s.t.\;\;\;\;\;\;{\mathbf{q}} \in {\mathds{Q} } \hfill \\
\end{gathered}
\end{equation}
where ${\mathds{Q } }$ is the discrete phase alphabet which is discussed detailedly in the next section.

The optimization problem of ($12$) is non-convex and NP-hard in general, whose optimal solution cannot be found in polynomial time. To tackle this problem, we consider the convex hull of the feasible points for each dimension, and relax ($12$) to a continuous QP subproblem. In the next section, we introduce a novel algorithm---Continuous Approximation Method (CAM) to  approximate to the nearest feasible vector ${{\mathbf{q}}_{opt}}$ for the optimization problem of ($12$).

\section{Proposed Algorithm}

To describe ${\mathds{Q } }$ in ($12$), we suppose an extreme situation of ($9$) with $\delta  = 2\pi $, i.e., the similarity parameter $\varepsilon  = 2$, where the SC vanishes and only the CMC is in effect. As shown in Fig. 1, the feasible region of $s(k)$ is a full circle with a constant radius on the complex plane $\mathbb{C}$. In the discrete phase case, we construct a constellation with $\Omega $ points on the circle, i.e., ${q_1}, \ldots ,{q_\rho}, \ldots ,{q_\Omega }$, as an example with $\Omega  = 16$ shown in Fig. 1, where the radian between any two adjacent points is given as
\begin{equation}\label{13}
\tau  = \frac{{2\pi }}{\Omega }.
\end{equation}
The constellation points are distributed over the whole continuous feasible region for each dimension. Thus, each sample of the continuous transmit waveform can be quantized to the finite feasible points which can be given as
\begin{equation}\label{14}
{q_\rho } = \exp (j\rho \tau )/\sqrt {{N_T}N}
\end{equation}
where $\rho  = 1, \ldots ,\Omega$. For notational convenience, let
\begin{equation}\label{15}
{\Gamma _\Omega } = \left\{ {\left. {{q_\rho } = \exp (j\rho \tau )/\sqrt {{N_T}N} } \right|\rho  = 1, \ldots ,\Omega } \right\}
\end{equation}
represent the discrete phase alphabet for each dimension. As shown in Fig. 2 (a), by connecting all the feasible points, a regular polygon is formulated in the tint area, which is the convex hull of ${\Gamma _\Omega }$. Further, let ${\mathbf{\Lambda }_\Omega }$ be the discrete phase alphabet for all dimensions, which is given as
\begin{equation}\label{15}
{{\mathbf{\Lambda }}_\Omega } = \underbrace {{\Gamma _\Omega } \odot  \cdots  \odot {\Gamma _\Omega } \odot  \cdots  \odot {\Gamma _\Omega }}_{{\text{the number of }}{\Gamma _\Omega }{\text{ is }}{N_T}N}
\end{equation}
where $ \odot $ denotes the Cartesian product. Obviously, ${\mathbf{q}} \in {\mathbf{\Lambda }_\Omega }$, $q(k) \in {\Gamma _\Omega }$.

However, for the general situation of ($9$) with $\delta $ less than $2\pi $, the CMC and the SC are both involved. We introduce a positive even integer parameter $\eta $ to indicate the similarity between ${\mathbf{q}}$ and ${{\mathbf{q}}_0}$, where ${{\mathbf{q}}_0}$ represents the quantized reference waveform. By using $\eta $, we rewrite ($8$) as follows
\begin{equation}\label{17}
\begin{gathered}
  {\gamma _k} = \arg {q_0}(k) - \eta \tau /2 \hfill \\
  \varphi  = \eta \tau  \hfill \\
\end{gathered}
\end{equation}
where $q_0(k)$ is the $k$-th entry of ${\mathbf{q}_0}$. For each dimension, $\eta $ stands for the number of feasible points around $q_0(k)$, which is obviously less than $\Omega $. According to the second equation in ($8$), the actual similarity tolerance $\varepsilon $ between ${\mathbf{q}}$ and ${{\mathbf{q}}_0}$  can be expressed as
\begin{equation}\label{18}
\varepsilon {\text{ = }}\sqrt {2[1 - cos(\varphi /2)]}.
\end{equation}

For instance, Fig. 2 (a) shows the feasible points similar to ${q_0}(k)$ in the case of $\eta  = 6$, which can be enumerated as a finite set
\begin{equation}\label{19}
{\Gamma _\eta }(k) = \{ {p_1}(k), \ldots {p_{\eta /2 + 1}}(k), \ldots ,{p_{\eta  + 1}}(k)\}
\end{equation}
where the angle of ${p_1}(k)$ is ${\gamma _k}$, ${p_{\eta/2  + 1}}(k)$ is ${q_0}(k)$, and ${p_{\eta  + 1}}(k)$ is the mirror constellation point of ${p_1}(k)$ with respect to ${q_0}(k)$. The convex hull of ${\Gamma _\eta }(k)$ is shown as a darker polygon marked by $\Theta (k)$. Noting that ${\Gamma _\eta }(k) \subseteq {\Gamma _\Omega }$, any given ${{p}_\mu }(k)$, $\mu  = 1, \ldots ,\eta  + 1$, can be expressed as
\begin{equation}\label{20}
{p_\mu }(k) = \exp \{ j[{\gamma _k} + (\mu  - 1)\tau ]\} /\sqrt {{N_T}N}.
\end{equation}
Furthermore, let
\begin{equation}\label{21}
{\mathbf{\Lambda }_\eta } =  {{\Gamma _\eta }(1) \odot  \cdots  \odot {\Gamma _\eta }(k) \odot  \cdots  \odot {\Gamma _\eta }({N_T}N)}
\end{equation}
represent the feasible vector of ${\mathbf{q}}$ for all dimensions. It is obvious that ${\mathbf{\Lambda }_\eta} \subseteq {\mathbf{\Lambda } _\Omega }$, and especially, when $\eta  = \Omega $, ${\mathbf{\Lambda }_\eta } = {\mathbf{\Lambda } _\Omega}$.

\begin{figure}[t]
\centering
\subfigure[]{
\includegraphics[width=0.7\columnwidth]{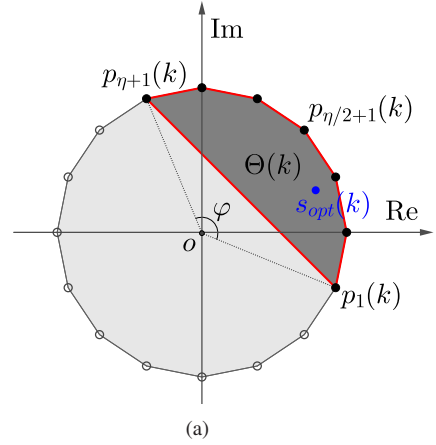}
}
\subfigure[]{
\includegraphics[width=0.8\columnwidth]{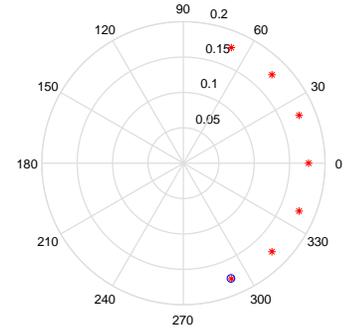}
}
\caption{(a) The convex hull of ${\Gamma _\Omega }$ with $\Omega = 16$ and the convex hull of ${\Gamma _\eta }(k)$ with $\eta = 6$ for $k$-th dimension; (b) A simulation example of the $k$-th entry of the optimal solution ${{\mathbf{s}}_{opt}}$ for the CAM approach.}
\label{2}
\end{figure}

We focus on the feasible points of ($12$) for each dimension, where we relax the finite feasible points set ${\Gamma _\eta }(k)$ to the convex hull itself, resulting in the following QP problem
\begin{equation}\label{22}
\begin{gathered}
  \mathop {\max }\limits_{\mathbf{s}} \;\;\; {{\mathbf{s}}^H}{\mathbf{Y}}{\mathbf{s}} \hfill \\
  s.t.\;\;\;\;\;\;{\mathbf{s}} \in \mathbf{\Delta}  \hfill \\
\end{gathered}
\end{equation}
where $\mathbf{\Delta}$ is the Cartesian product of the convex hull for all dimensions, which can be expressed as
\begin{equation}\label{23}
\mathbf{\Delta}  = \Theta (1) \odot  \cdots  \odot \Theta (k) \odot  \cdots  \odot \Theta ({N_T}N).
\end{equation}
Obviously, the optimization problem of ($22$) is convex, which can be solved via numerical solvers, e.g., the CVX toolbox. As shown in Fig. 2 (a), ${s_{opt}}(k)$ is the $k$-th entry of the optimal solution ${{\mathbf{s}}_{opt}}$ of ($22$). In Fig. 2 (b), we take a simulation example of ${s_{opt}}(k)$ for the CAM approach with $\Omega = 16$ and $\eta = 6$. The red stars indicate the feasible points. The blue circle is ${s_{opt}}(k)$, which obviously locates in the convex hull.

We provide the analytical formulas of the convex hull $\Theta (k)$ for each dimension in the sequel. For notational convenience, we omit the dimension number $k$. As shown in Fig. 2 (a), the convex hull has $\eta  + 1$ edges which are highlighted in red, i.e., $L_1: {p_1}{p_2}, \ldots , L_\mu: {p_\mu }{p_{\mu  + 1}}, \ldots , L_\eta: {p_\eta }{p_{\eta  + 1}}$, and $L_{\eta+1}: {p_{\eta  + 1}}{p_1}$. Let us define the middle point of ${p_\mu }{p_{\mu  + 1}}$ as
\begin{equation}\label{24}
{m_\mu } = \frac{{{p_\mu } + {p_{\mu  + 1}}}}{2}, \mu  = 1, \ldots , \eta.
\end{equation}
According to the basic plane analytic geometry, given any $s \in \mathbb{C}$, the line ${L_\mu }$ can be expressed as
\begin{equation}\label{25}
\text{Line }{L_\mu }:{f_\mu }(s) = Re(m_\mu ^ * (s - {m_\mu })) = 0.
\end{equation}
In addition, for the $(\eta +1)$-th edge, the middle point ${m_{\eta+1} }$ is
\begin{equation}\label{26}
{m_{\eta+1} } = \frac{{{p_{\eta+1} } + {p_1}}}{2},
\end{equation}
and the corresponding line ${L_{\eta+1} }$ is given as
\begin{equation}\label{27}
\text{Line }{L_{\eta+1} }:{f_{\eta+1} }(s) = Re(m_{\eta+1} ^ * (s - {m_{\eta+1} })) = 0.
\end{equation}
As a consequence, the $k$-th convex hull for each dimension can be separately formulated in two cases. For the first one, when $0 \le \varphi < \pi $, namely $0 \le \eta < \Omega/2$ $, \Theta (k)$ is given by
\begin{equation}\label{28}
\left\{ {\begin{array}{*{20}{c}}
  {{f_\mu }(s) \le 0,\mu  = 1, \ldots ,\eta } \hfill \\
  {{f_{\eta  + 1}}(s) \ge 0}. \hfill \\
\end{array}} \right.
\end{equation}
For the case of $\pi \le \varphi \le 2\pi $, namely $\Omega/2 \le \eta \le \Omega$, $\Theta (k)$ is given by
\begin{equation}\label{29}
\left\{ {\begin{array}{*{20}{c}}
  {{f_\mu }(s) \le 0,\mu  = 1, \ldots ,\eta } \hfill \\
  {{f_{\eta  + 1}}(s) \le 0}. \hfill \\
\end{array}} \right.
\end{equation}

Finally, by comparing the Euclidean distance between ${s_{opt}}(k)$ and the feasible points in ${\Gamma _\eta }(k)$, the feasible point with the minimal distance is checked for each dimension. Thus, we obtain the optimal feasible vector ${{\mathbf{q}}_{opt}}$.

\emph{Remark}: The complexity of the CAM mainly comes from computation of solving the QP problem. By using the interior-point algorithm, the total number of iterations needed is $\mathcal{O}(\sqrt {{N_T}N}\log (1/\varepsilon ) )$, and each iteration can be executed in ${\mathcal{O}}({N}_T^{3}{{N}^{3}})$ arithmetic operations \cite{Monteiro1989}. Meanwhile, considering the finite discrete alphabet, the optimization problem of ($12$) can be also solved by the exhaustive search method, while the worst case complexity of such method is exponential.

\section{Numerical Results}

\begin{figure*}[t]
\centering
\subfigure[]{
\includegraphics[width=0.62\columnwidth]{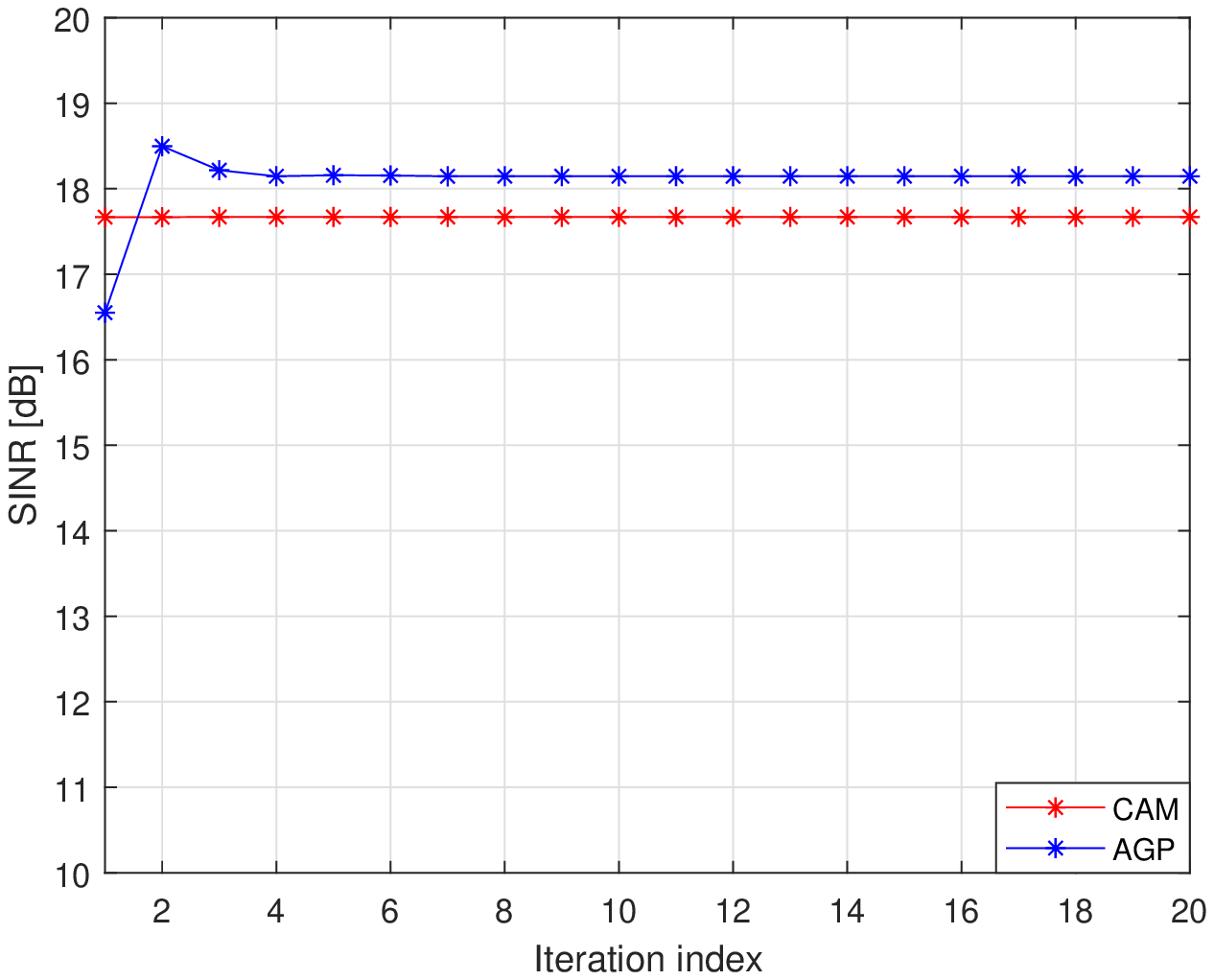}
}
\hspace{0cm}
\subfigure[]{
\includegraphics[width=0.62\columnwidth]{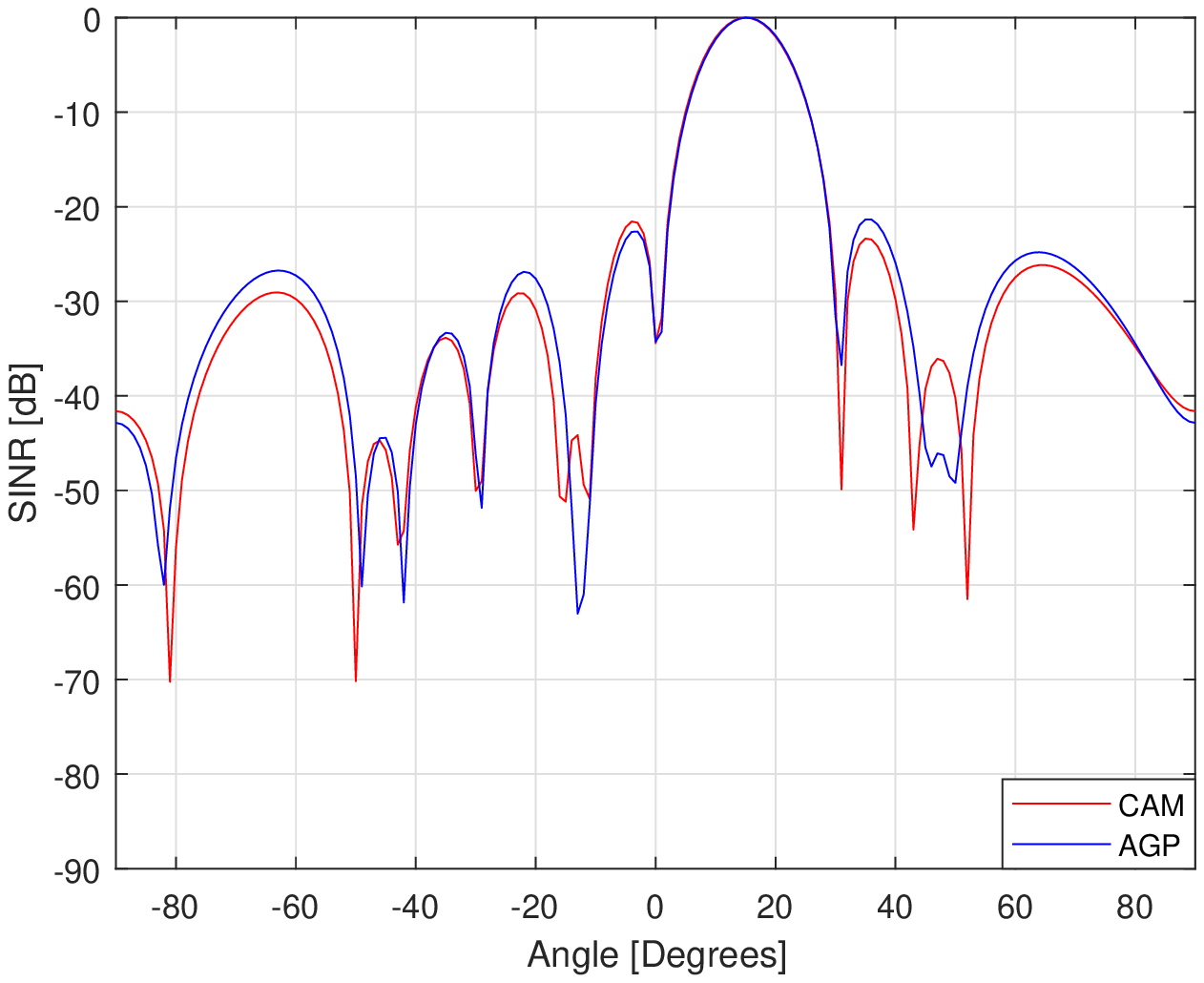}
}
\hspace{0cm}
\subfigure[]{
\includegraphics[width=0.62\columnwidth]{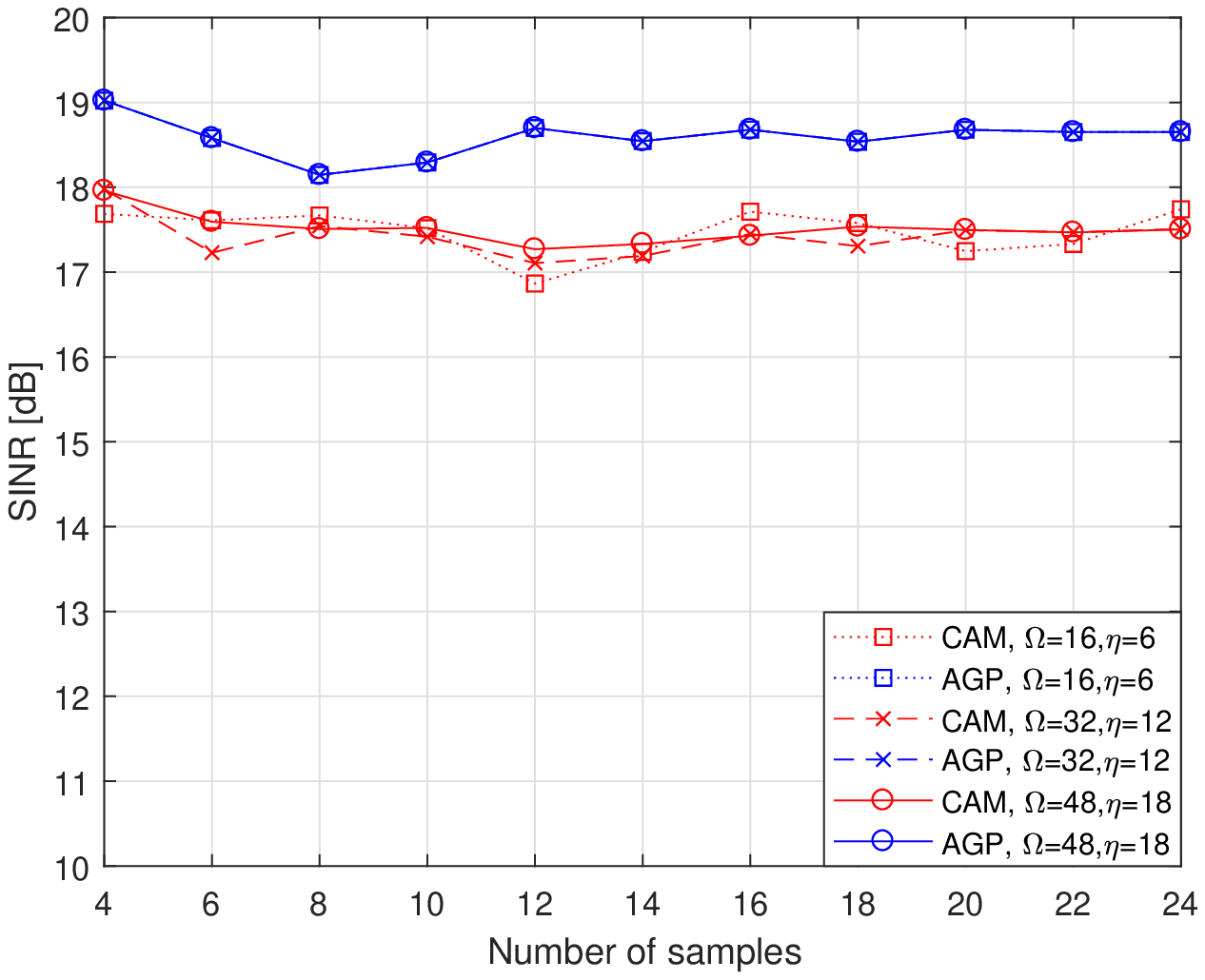}
}
\caption{Performance comparison between the CAM in discrete phase case and the AGP in continuous phase case: (a) SINR in each iteration; (b) Beampattern; (c) SINR with increasing number of samples.}
\label{3}
\end{figure*}
In this section, we provide numerical results to evaluate the performance in terms of the SINR and the beampattern between the CAM in discrete phase case and the AGP \cite{arxiv1805} in continuous phase case, and use chirp waveform as our benchmark. For the continuous phase case, the reference waveform ${{\mathbf{s}}_0} \in \mathbb{C}^{{N_T}N \times 1}$ can be obtained by stacking the columns of ${{\mathbf{S}}_0}$, which we choose as the following orthogonal chirp waveform matrix without loss of generality
\begin{equation}\label{29}
{{{S}}_0}(k,n) = \frac{{\exp [j2\pi k(n - 1)/N]\exp[j\pi {{(n - 1)}^2}/N]}}{{\sqrt {{N_T}N} }}
\end{equation}
where $k = 1,...,{N_T}$, $n = 1,...,N$. For the discrete phase case, the reference waveform ${{\mathbf{q}}_0}$ is obtained by quantizing ${{\mathbf{s}}_0}$. The numbers of the transmit and the receive antennas are ${{{N}}_T} = 4$, ${{{N}}_R} = 8$, respectively. In addition, we consider a scenario with three fixed signal-dependent clutters and additive white Gaussian disturbance with variance ${\sigma _n} = 0$dB. The target is located at an angle ${\theta _0} = {15^ \circ }$ with a reflecting power of ${\left| {{\alpha _0}} \right|^2} = 10$dB and three fixed interference sources located at ${\theta _1} = {-50^ \circ }$, ${\theta _2} = {-10^ \circ }$ and ${\theta _3} = {40^ \circ }$ reflecting a power of ${\left| {{\alpha _1}} \right|^2} = {\left| {{\alpha _2}} \right|^2} = {\left| {{\alpha _3}} \right|^2} = 30$dB.

As shown in Fig. 3 (a), we firstly compare the SINR in each iteration between the CAM and the AGP with $N = 8$, $\Omega = 16$ and $\eta = 6$. According to ($13$), ($17$) and ($18$), we have $\varphi = 3\pi/4$ and $\varepsilon \approx 1.1$. The SINR resulting from the CAM is slightly inferior to the AGP for the continuous phase case. Fig. 3 (b) shows the beampattern of the optimal waveform for both methods, using the same parameters as in Fig. 3 (a). The simulation reveals that the CAM achieves a comparable beampattern performance with that of the continuous phase case.

In Fig. 3 (c), we compare the SINR for both methods with increasing number of samples $N$. In order to normalizing the actual similarity tolerance $\varepsilon$, we change $\Omega$ and $\eta$ with a fixed ratio 8/3, i.e., $\Omega = 16$ and $\eta = 6$, $\Omega = 32$ and $\eta = 12$, $\Omega = 48$ and $\eta = 18$, leading to the same $\varepsilon$ as Fig. 3 (a). The AGP results in a consistent SINR with different $\Omega$ and $\eta$. Meanwhile, the SINR of the CAM exhibits an acceptable fluctuation, which is only 1dB lower than its AGP counterpart substantially.

\section{Conclusion}
In this letter, we propose a novel approach for MIMO radar waveform design with constant modulus and discrete phases. By constructing the polygon convex hull of the feasible constellation points, we develop a CAM algorithm to relax the discrete optimization problem as sequential convex QP subproblems. The optimal solution is then obtained by quantizing the solution of QP to its nearest neighbor. Compared with the exhaustive search, the computational complexity has been significantly reduced. Numerical results reveal that the proposed approach only yields slight performance-loss with respect to that of the continuous phase case.


%

%


\ifCLASSOPTIONcaptionsoff
  \newpage
\fi



\bibliographystyle{IEEEtran}
\bibliography{IEEEabrv,Ren_REF}
\end{document}